\documentclass[a4paper,floatfix,rmp,twocolumn,showkeys,superscriptaddress]{revtex4}
\usepackage[latin1]{inputenc}
\usepackage{graphicx}

\begin{document}

\title{Synthetic biocomputation design using supervised gene regulatory networks}

\providecommand{\ICREA}{ICREA-Complex Systems Lab, Universitat Pompeu Fabra, Dr Aiguader 88, 08003 Barcelona, Spain}
\providecommand{\SFI}{Santa Fe Institute, 1399 Hyde Park Road, Santa Fe NM 87501, USA}
\providecommand{\IBE}{Institut de Biologia Evolutiva, UPF-CSIC, Psg Barceloneta 37, 08003 Barcelona, Spain}

\author{Lu\'is F. Seoane\footnote{corresponding author}} \affiliation{\ICREA} \affiliation{\IBE}
\author{Ricard V. Sol\'e\footnote{corresponding author}}   \affiliation{\ICREA}\affiliation{\IBE}  \affiliation{\SFI}

	\begin{abstract}

		The potential of synthetic biology techniques for designing complex cellular circuits able to
solve complicated computations opens a whole domain of exploration, beyond experiments and theory.
Such cellular circuits could be used to carry out hard tasks involving decision-making, storage of
information, or signal processing. Since Gene Regulatory Networks (GRNs) are the best known
technical approach to synthetic designs, it would be desirable to know in advance the potential of
such circuits in performing tasks and how classical approximations dealing with neural networks can
be translated into GRNs. In this paper such a potential is analyzed. Here we show that feed-forward
GRNs are capable of performing classic machine intelligence tasks. Therefore, two important
milestones in the success of Artificial Neural Networks are reached for models of GRNs based on Hill
equations, namely the back-propagation algorithm and the proof that GRNs can approximate arbitrary
positive functions. Potential extensions and implications for synthetic designs are outlined.

	\end{abstract}


\maketitle

	\section{Introduction}
		\label{sec:intro}

		Cells are entangled living machines capable of very complex computational tasks. They rival
parallel computers and are in charge of the fine tuned responses that allow them, along with tissues
and organs, to properly adapt to external and internal challenges \cite{Bray1990, Bray1995}. The
idea that complex patterns of cellular behavior can be an emergent property of	gene-gene
interactions was early proposed by Stuart Kauffman, who used a Boolean approximation to gene
regulation as a minimal model of the true (and complex) molecular events \cite{Kauffman}. This view
was originated shortly after the classical work by Warren McCulloch and Walter Pitts, which showed
that any particular computational task (as defined by a logic gate) can be mapped into a 
threshold-like neural network \cite{McCulloch-Pitts}. It is interesting to notice that, since those
early years, both neural and genetic networks have received an always increasing attention both at
the level of the details of their interacting constituents as well as in terms of theoretical
models. In both cases, it is often possible to consider that each element (either a formal neuron or
a simplified gene, which we shall call indistinctly \emph{unit} throughout the text) responds to
external stimuli in nonlinear ways. The associated response functions that characterize both types
of units are commonly stepwise, ideally Boolean-like.

		The theory of neural networks rapidly advanced, primarily thanks to the development of computers
and simulation techniques. In parallel, neuroscience actively studied the behavior of nerve cells
with outstanding precision. At the end of the 20th century, neuron-based models were highly accurate
and essentially well established \cite{Dayan-Abbott}. Moreover, several standard approximations
emerged towards the distributed solution of a plethora of computation problems \cite{Bishop}:
leaving aside elemental classification and interpolation tasks based on sample data; Artificial
Neural Networks (ANNs) were successfully applied on signal processing, pattern recognition, complex
inference, nonlinear control, etc. By contrast, gene networks received less consideration since
molecular biology started to dominate the scene from the 1950s. More and more attention was being
paid to how single genes worked and a largely reductionistic agenda was developed. With the
discovery of gene regulation by Jacob and Monod \cite{Jason-Monod}, the picture started to (very)
slowly move towards a circuit-based functional view of cellular control. The gene network view has
been ever since gradually adopted by most biologists, who became aware of the dominant role played
by gene-gene interactions.

\begin{figure*}
			\includegraphics[width=0.85\textwidth]{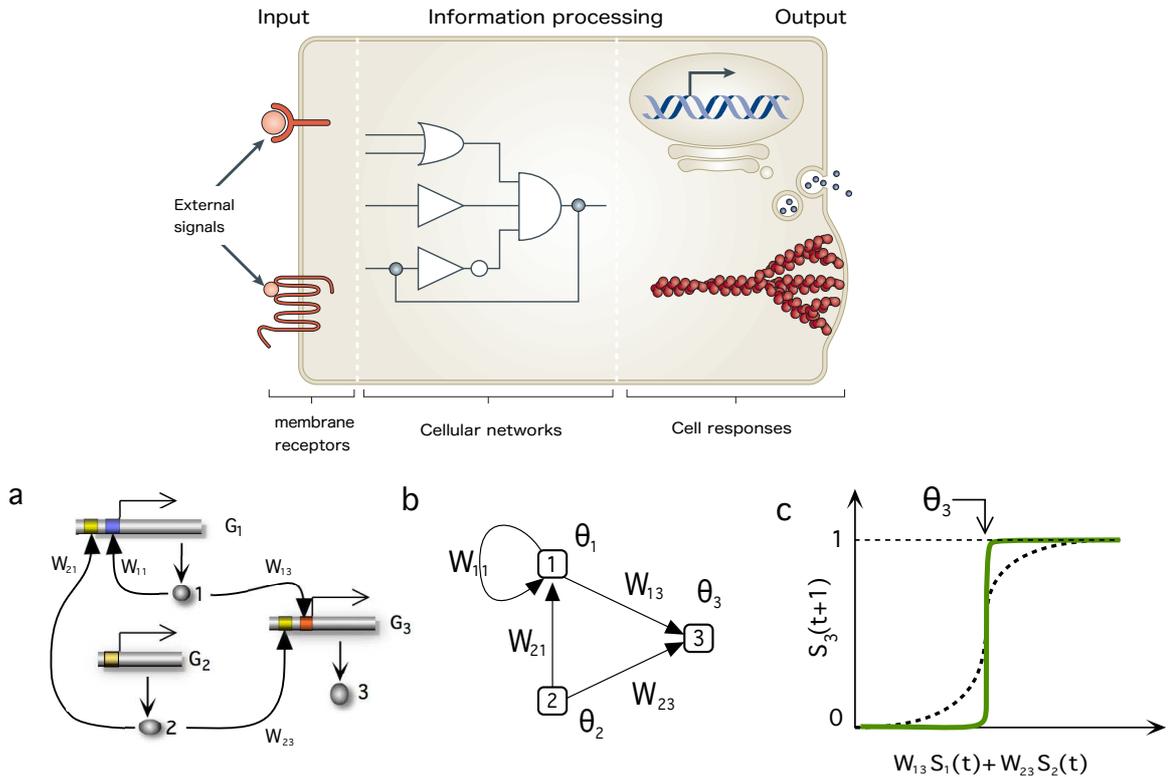}

			\caption{\textbf{Information processing in living cells} Cells are complex living 
			computational devices. The upper diagram (picture adapted from Lim 2010) is a simplified drawing including some basic components of the computational logic of cells. In this simplified description, cells gather signals 
			of different nature from the external world (while sensing their internal state) and respond to 
			these signals by means of information-processing networks, eventually triggering 
			biochemical and physical changes as output. \textbf{a} Gene networks are one 
			type of such molecular machinery. Here we represent their expression and response 
			to DNA-binding proteins (so called transcription factors) using arrows to indicate 
			who influences whom and weights $W_{ij}$ to represent the strength of the regulatory 
			interaction. A formal approach to the same network is shown in  \textbf{b} 
			where genes are formally represented as state variables in a nonlinear dynamical system (see text). 
			The responses are often sharp, reminding us the ones observed in neural 
			systems. In  \textbf{c} an example of the type of 
			nonlinearity considered here is shown. Here the activation 
			response of a given element ($S_3$) requires a total input reaching a value higher than a critical 
			threshold $\theta_3$. Idealized models consider a all-or-none response (red line) but 
			real systems follow a smooth profile (dashed line) that can be characterized by means of Hill functions. }
			\label{fig:intro.01}
		\end{figure*}

    Gene Regulatory Networks (GRNs) are a holistic conceptualization of large assemblies of
interacting genes and their regulatory interplay \cite{Kitano, bookAlon}. They are the result of a non-designed 
process of evolutionary tinkering \cite{CRCaso} and thus they display some non-standard 
patterns of network organization. Specific mathematical
modeling can be made thanks to the tools already developed for single gene and other molecular
dynamics characterization: we will be working with the quite standard and successful Hill
differential equation \cite{Hill}. Reconstruction of small motifs from real genetic networks
\cite{Alon}, and also of each time larger interwoven collections of genes \cite
{SaezRodriguez-Sorger} has become possible using experimental data. Small feed-forward motifs like
those represented in figure \ref{fig:intro.01} are overly abundant in large GRNs \cite
{Macia-Sole2009a, Widder-Macia}, and they resemble us enough of the synthetic networks that we will
be designing. The connection that we intend to make between GRNs and machine learning is already
hinted at by AI methods for gene-webs reconstruction \cite{Sirbu-Crane}, where ANNs reveal
themselves as a very appropriate model of GRNs.

    The understanding of how GRNs work helps us build a much better picture of how computations
occur in living cells \cite{Bray1990, Bray1995, Nurse, Brenner} and invites us to think of GRNs from
an engineering standpoint. In recent years ambitious calls and claims have been made that pursue the
implementation of actual biological computing devices \cite{Amos2004, Amos2008, Benenson}. These
would use interconnected genes, assemblies of cells, and alike to reckon and execute controlled
responses to external conditions. One final goal would be to elaborate computing resources that
could be easily integrated into living organisms, although up to date most of these contributions
have been developed in vitro. Even in an artificial environment free from the unpredictability of
complex organisms, the components that nature provides us with present a series of important
drawbacks: our theoretical understanding of cellular processes is largely incomplete; we usually
handle leaky, diffusing systems where a precise spatial architecture becomes difficult, if not
impossible; chemical reactions are essentially stochastic; etc. Thus, seemingly simple designs have
required very rigorous experimental controls \cite{Rothemund-Winfree, Bratsun-Hasty, 
Tabor-Ellington, Mondragon-Hasty}. Major progresses might require non-standard approximations 
\cite{Regot-Sole, Macia-Sole2012} that liberate us from any of the existing constrains, or to
novel technical advances, such as multiples genome engineering \cite{Wang-Church}. The exhaustive experimental \cite{Hoffman-Kipp} or theoretical \cite {Hasty-Collins} characterization of specific existing circuits (both in vitro and 
in silico) is also necessary.

    In this paper we tackle theoretical issues regarding the most basic designable computational
capabilities of GRNs. We restrict ourselves to feed-forward networks, which lack feedback loops and
thus will not exhibit dynamical behaviors other than steady state attractors. We present an error
back-propagation algorithm to design actual GRNs that would solve specific problems. Experimentally
hard wiring a set of parameters into a synthetic biological system seems a little ahead of schedule, 
but the development of genetic engineering techniques that allow a combinatorial design of complex circuits is 
becoming a reality \cite
{Mattiussi-Floreano, Wang-Church} and we shall soon have the potential of creating any requested
synthetic GRN. Similarly, synthetic customizable signaling networks are becoming a reality \cite
{Bashor-Lim, Lim}. Although combinatorial design with biological components is still in its infancy,
the potential for complex computational synthetic networks is growing fast \cite{Sole-Macia}. The
questions we pose are: How do these gene-based circuits need to be designed and to what extent can
they perform computational tasks similar to those performed by ANNs? Are these tasks easy to
implement by GRN? What are the design rules required to obtain the optimal designs?

		We close this introduction outlining that GRNs are just one of the many biological structures
that seem an appropriate substrate to implement AI means. We can think of others, such as
transduction networks, that can also be modeled up to a great detail using Hill equations. Hill
functions will be our main tool in the current paper, thus our developments should apply not only to
GRNs, but to many more systems. \\

		The article is structured as follows: In section \ref{sec:proof} it is shown in a constructive
way how a feed-forward network of Hill equations can approximate any arbitrary continuous positive
function. In section \ref{sec:back} the back-propagation rules are explicitly derived for our
mathematical model of GRNs and in section \ref{sec:ex} we show how our idea was applied (of course,
in silico) to one fitting problem and three classification tasks. Discussion and future lines of
work follow in section \ref{sec:disc}.

	\section{Function approximation by synthetic GRN}
		\label{sec:proof}

		In this section we address the demonstration that finite feed-forward networks of Hill equations
can approximate any positive, bounded, continuous function defined over a bounded subset of the real
numbers: such systems already incorporate the needed non-linearities that linear perceptrons were
missing. For the demonstration we use only Hill equations at their steady state, which already
encompass enough complexity for our purposes. This work shall just be a glance into the actual
capabilities of GRNs if we would use them as computing devices. Incorporating feedback or analyzing
the evolution in time would reveal GRNs as appropriate substrates to biologically implement more
powerful devices such as recurrent or echo state networks \cite{Maass-Markram}.

		\subsection{Mathematical characterization of Hill functions}
			\label{subsec:characterization}
			
      Hill differential equations model the temporal dynamics of the concentration $y(t)$ of a
protein $Y$ regulated by a set of $N$ promoter proteins $X_i$ whose concentrations are $x_i(t)$. In
the literature we can find different implementations of these dynamics depending on whether or not
different promoters can associate with each other to express $Y$ \cite{Hasty-Collins, 
Goutelle-Maire, Santillan}. We choose a formalism in which such cooperation is banned:
				\begin{eqnarray}
					{d y\over dt} &=& {\alpha_0 + \alpha_1 x_1^{n_1} + ... + \alpha_N x_N^{n_N}\over 
								\beta_0 + \beta_1 x_1^{n_1} + ... + \beta_Nx_N^{n_N}} -\kappa y. 
					\label{eq:proof.01}
				\end{eqnarray}
Here $\kappa$ represents the degradation rate of $Y$, $n_i$ with $i=1,...,N$ are Hill coefficients
that estimate the number of $X_{i}$ molecules required for a functional effect on $Y$, $\alpha_i$
and $\beta_i$ indicate the affinity of $Y$ to each one of the regulating proteins, and $\alpha_0$
and $\beta_0$ encode for the basal activity of $Y$ (i.e. the concentration of that protein when none
of the regulating agents is present).

			Other formulations of the sought dynamics are qualitatively similar regarding the task that we
have ahead. If $Y$ is regulated by just one promoter $X$ with concentration $x$, equation
\ref{eq:proof.01} reduces to:
				\begin{eqnarray}
					{d y \over dt} &=& {\alpha_0 + \alpha_1 x^n \over 
							\beta_0 + \beta_1 x^n } - \kappa y. 
					\label{eq:proof.02}
				\end{eqnarray}
Let us note that a modeling allowing a cooperative action of different $X_i$ upon $Y$ also reduces
to this expression for just one regulating protein, thus the following results are general.

      We will see now how a clever use of such a simple motif is enough to show the very rich
computational capabilities of networks of GRNs. We begin our study by a thorough characterization of
the steady state of this equation, part of which might already be found in the literature. \\

			For a fixed concentration $x$ of protein $X$, equation (\ref{eq:proof.02}) decays towards:
				\begin{eqnarray}
					\tilde{y}(x) = {\alpha_0 + \alpha_1 x^n \over 
								\kappa\left[\beta_0 + \beta_1 x^n\right] }, 
					\label{eq:proof.03}
				\end{eqnarray}
$\tilde{y}(x)$ representing the concentration of $Y$ at the steady state as a function of the
concentration of its promoter. The slope of this function with respect to $x$ reads:
				\begin{eqnarray}
					{d \tilde{y}(x)\over dx} &=& {n(\alpha_1\beta_0-\beta_1\alpha_0) x^{n-1}\over 
								\kappa \left[\beta_0+\beta_1x^n\right]^2} 
								\equiv D_{\tilde{y}}(x). 
					\label{eq:proof.04}
				\end{eqnarray}
Let us note that $(\alpha_1\beta_0-\beta_1\alpha_0)$ determines the sign of this slope and that this
sign remains unchanged for the whole domain of $\tilde{y}(x)$, meaning that this function is either
monotonously increasing or decreasing.

			If we calculate the second derivative of $\tilde{y}(x)$: 
				\begin{eqnarray}
					{d D_{\tilde{y}}(x)\over dx} \equiv {d^2 \tilde{y}(x)\over dx^2} &=& n(\alpha_1\beta_0-\beta_1\alpha_0)x^{n-2} \nonumber \\
								&&\times {(n-1)\beta_0 - (n+1)\beta_1 x^n \over \kappa \left[ \beta_0+\beta_1 x^n \right]^3}, 
					\label{eq:proof.05}
				\end{eqnarray}
and calculate the extrema of $D_{\tilde{y}}(x)$, ${d D_{\tilde{y}}(x)\over dx}=0$; we find three
different solutions:
				\begin{itemize}
					\item $x^{n-2}=0 \iff x=0\equiv x_0, \> n>2$. 
					\item ${1\over \left[ \beta_0+\beta_1 x^n \right]^3}\rightarrow 0 
							\iff x\rightarrow +\infty \equiv x_{\infty}$. 
					\item $(n-1)\beta_0 - (n+1)\beta_1 x^n = 0 \\ 
							\iff x = \left[{(n-1)\beta_0 \over (n+1)\beta_1}\right]^{1/n} 
							\equiv x_\theta$. 
				\end{itemize}

	\begin{figure*}
		\includegraphics[width = 15 cm]{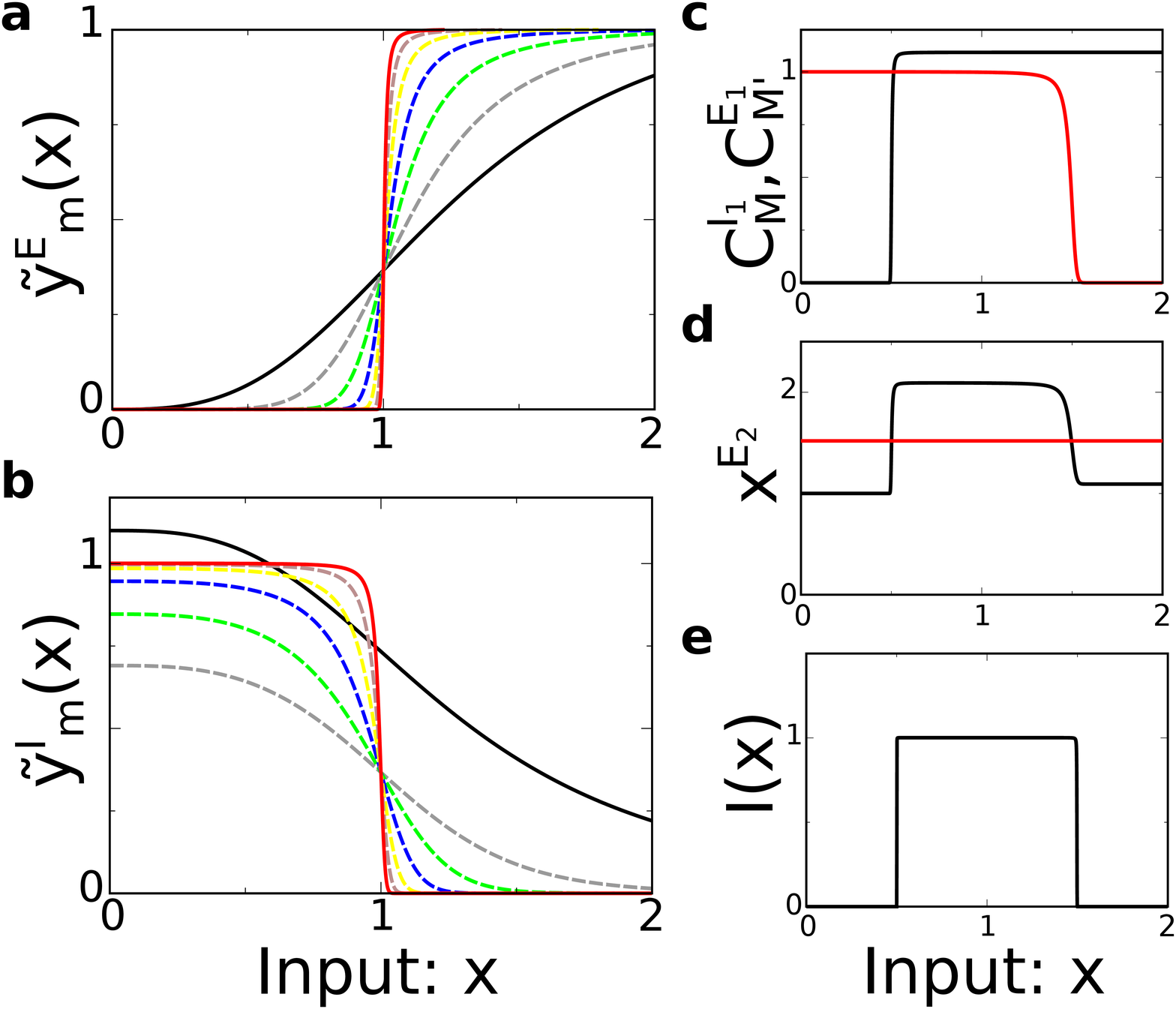}

    \caption{\textbf{Approximating a rectangular function with superpositions of Hill equations. }
\textbf{a} A superposition of excitatory Hill equations $\tilde{y}^E_m(x)$, with $m=1,...,7$ builds
up an  excitatory cascade $C^E_7(x)$ that approaches an excitatory step function. \textbf{b} An
inhibitory Hill equation  $\tilde{y}^I_1(x)$ is combined with a superposition of excitatory Hill
equations $\tilde{y}^E_m(x)$, with  $m=2,...,7$ to build up an inhibitory cascade $C^E_7(x)$ that
approaches an inhibitory step function.  $\tilde{y}^E_1(x)$ in \textbf{a} and $\tilde{y}^I_1(x)$ in
\textbf{b} are plotted with a continuous black line,  and in both \textbf{a} and \textbf{b}
$\tilde{y}^E_7(x)$ is plotted with a continuous red line. Intermediate  equations
($\tilde{y}^E_m(x)$; $m=2,...,6$) are always plotted with dashed lines. Steepness grows with $m$.
The upper limit of the Hill equations has been set such that all excitatory units take values
approximately  between $0$ and $1$. Therefore, the inhibitory unit in \textbf{b} had to be designed
with a $\tilde{y}^I_1(0)>1$,  $0<\tilde{y}^I_1(\infty)<\tilde{y}^I_1(0)$; so it is not an strictly
excitatory unit as they were defined  on the text.
    \textbf{c} Excitatory $C^{E_1}_{M'}(x)$ (black) and inhibitory $C^{I_1}_{M}(x)$ (red) cascades
fulfilling  the necessary relationships to build a rectangular function. Back to the biological
interpretation of our model,  we must assume that both cascades are different metabolic pathways
promoted by a same protein $X$ that presents a  concentration $x$ working as the input of the whole
system. Both pathways would produce a same protein $X^{E_2}$,  whose joint yield as a function of
$x$ is shown in panel \textbf{d}. A third cascade must be added, $C^{E_2}_{M''}(x)$:  an excitatory
one, and its threshold (horizontal red line in panel \textbf{d}) must be appropriately chosen to
approach a rectangular function $R(x)$. \textbf{e} Tuning $C^{E_2}_{M''}(\infty)$ we can approximate
the desired  $R(x)$: A rectangular function with $R(x)=1$ if $x\in[0.5, 1.5)$ and $0$ elsewhere is
faithfully approximated by  $I(x)\equiv C^{E_2}_{M''}(x)$ (black). }

		\label{fig:proof.01}
	\end{figure*}

			The last extremum of $D_{\tilde{y}}(x)$ is the most interesting one as it indicates an
inflexion point of the original function $\tilde{y}(x)$. For convenience and similarity to sigmoids,
we will call $x=x_\theta$ the threshold of the Hill equation -- thus the notation. We evaluate the
slope at the threshold:
				\begin{eqnarray}
					D_{\tilde{y}}(x_\theta) &=& {(n+1)^2 (\alpha_1\beta_0-\beta_1\alpha_0) \over 4 n\kappa \beta_0^2 }
									\left[ {(n-1)\beta_0 \over (n+1)\beta_1}\right]^{{n-1\over n}}
					\label{eq:proof.06}
				\end{eqnarray}\\

			By now we have got our Hill equation characterized by parameters that are rather abstract from
a geometric point of view, although their biological meaning is clear. Let us write $\alpha_0$,
$\alpha_1$, $\beta_0$, and $\beta_1$ in terms of $\tilde{y}(0)$, $\tilde{y}(\infty)$, and
$x_\theta$; which have got a more intuitive geometric interpretation. For the upper and lower limits
of $\tilde{y}(x)$ we get:
				\begin{eqnarray}
					\lim_{x \to 0^+} \tilde{y}(x) &=& {\alpha_0 \over \kappa \beta_0} \equiv \tilde{y}(0) \Rightarrow 
											\alpha_0 = \kappa \beta_0 \tilde{y}(0). \nonumber \\
					\lim_{x \to +\infty} \tilde{y}(x) &=& {\alpha_1 \over \kappa \beta_1} \equiv \tilde{y}(\infty) \Rightarrow
											\alpha_1 = \kappa \beta_1 \tilde{y}(\infty). 
					\label{eq:proof.07}
				\end{eqnarray}
From the equation for $x_\theta$ we can work out the ratio $\beta_0/\beta_1$:
				\begin{eqnarray}
					{\beta_0 \over \beta_1} &=& {n+1\over n-1} x_\theta^n, 
					\label{eq:proof.08}
				\end{eqnarray}
and substituting in equation (\ref{eq:proof.06}) we can write the slope of $\tilde{y}(x)$ at the
threshold as a function of the desired parameters:
				\begin{eqnarray}
					D_{\tilde{y}}(x_\theta) &=& \left[\tilde{y}(\infty)-\tilde{y}(0)\right]{(n-1)(n+1)\over 4n}x_\theta^{n-2}. 
					\label{eq:proof.10}
				\end{eqnarray}
		
      It is clear that $\tilde{y}(0)$, $\tilde{y}(\infty)$, and $x_\theta$ should be given if we
wanted to build up a Hill function with customized upper and lower bounds and threshold. It would
only remain $n$ to control the steepness given by equation \ref{eq:proof.10}. Indeed:
				\begin{eqnarray}
					\lim_{n \to 1^+} \|D_{\tilde{y}}(x_\theta)\| &=& 0, \nonumber \\ 
					\lim_{n \to +\infty} \|D_{\tilde{y}}(x_\theta)\| &=& +\infty; 
					\label{eq:proof.11}
				\end{eqnarray}
while $\left[\tilde{y}(\infty)-\tilde{y}(0)\right]$ still controls the sign of the bare derivative. 
This indicates that varying $n$ we can design Hill equations that are more or less flat. This also 
makes manifest the role of a large $n$ to introduce non-linearities in protein regulation processes, 
as well as its importance for sharply triggered dynamics. 
		
			Let us note that we went from a parameterization using $\kappa$, $n$, $\alpha_0$, $\alpha_1$,
$\beta_0$, and $\beta_1$ to one that only uses $n$, $\tilde{y}(0)$, $\tilde{y}(\infty)$, and
$x_\theta$. This is possible because $\kappa$ plays a normalizing role and can be absorbed into some
other constant. Also, we saw that $\beta_0$ and $\beta_1$ only affect the shape of the Hill
equations through their ratio.

			Now, by choosing $\tilde{y}(0)$ and $\tilde{y}(\infty)$ we can build a Hill function with any
wished upper and lower bounds, and choosing $x_\theta$ and $n$ we can set up any desired threshold
and steepness. From these choices we could work out the values of the original set of parameters and
we would still have freedom to choose $\kappa$ and either $\beta_0$ or $\beta_1$.

      A key ingredient in our demonstration will be that we can approximate a step function with any
desired accuracy. This can be done taking $n\rightarrow \infty$, and this would be completely fair
from a mathematical point of view. However, our endeavor is to show the computational capacity of
realistic biological systems. Regulatory systems such as those described by equation
\ref{eq:proof.02} usually present low values of $n$. A typical value is $n\simeq2$, which is
consistent with molecular processes based on dimerization \cite{bookAlon, Macia-Sole2009b}. Luckily
enough, there is a very interesting way around to approximate a step function by combining Hill
equations; and one that the nature itself seems to have used prominently for threshold-dynamics
regulation. This is explored in the following subsection.

		\subsection{Cascades of Hill equations}
			\label{subsec:cascades}
			
			An interesting feature of sigmoid functions is that they map an interval (let us say
$x\in(0,+\infty)$) into another (say $y\in(0,+1)$) in an exponential fashion, meaning that a linear
increase in $x$ corresponds to a somehow exponential modification of $y$. This can be seen in the
exponential decay of $y$ towards $0$ or $1$ when $x\rightarrow 0$ or $x\rightarrow +\infty$ and in
the exponential increase of $y(x)$ when $x$ approaches the threshold of the sigmoid. If this
characteristic is also true for Hill functions we can use the output of one such equation as the
input of another with the hope of transiting exponentially faster through the threshold of the
second equation when varying the original variable, thus providing a steepest overall dependency.
This turns out to be the case as it is demonstrated in figures \ref{fig:proof.01}\textbf{a} and
\ref{fig:proof.01}\textbf{b}.

			Using combinations of Hill equations, in this subsection we aim directly at building a step
function and a piecewise defined function that is constant and different from $0$ for a range of
$x$, and $0$ out of this range -- i.e. a rectangular function $R(x)$. In doing so we will disregard
other interesting functions that combinations of Hill equations might be producing. In this line, we
begin by calling strictly \emph{excitatory} ($\tilde{y}^E(x)$) and strictly \emph{inhibitory}
($\tilde{y}^I(x)$) Hill equations to those with $\tilde{y}^E(0)=0$, $\tilde{y}^E(\infty)>0$ and
$\tilde{y}^I(0)>0$, $\tilde{y}^I(\infty)=0$ respectively. 

			We define a \emph{$M$-Cascade function} -- and note it $C_M(x)$ -- as the system of $M$ Hill
equations coupled such that:
				\begin{eqnarray}
					C_M(x) &=& \tilde{y}_M\left(C_{M-1}(x)\right) , \nonumber \\ 
					\vdots & \vdots & \vdots \nonumber \\
					C_1(x) &=& \tilde{y}_1(x), 
					\label{eq:proof.12}
				\end{eqnarray}
This system can have very rich dynamics depending on the set $\{\tilde{y}_m(x); m=1,...,M\}$ of Hill
equations used, which might be inhibitory or excitatory; but for our immediate purpose (approaching
a step function) we are just interested in a set of equations such that all of them transit through
their threshold at the same time, thus making sure that $C_M(x)$ has got a well defined threshold
$x^{C_M}_{\theta}$ itself. To do this, we must build the $\tilde{y}_m(x)$ such that $x_{\theta,m} =
\tilde{y}_{m-1}(x_{\theta,m-1})$ for all $m$ except $x_{\theta,1}\equiv x^{C_M}_{\theta}$, which
still remains free for us to choose. This is always possible for any $x^{C_M}_{\theta}>0$, as we saw
in the previous subsection.

			We can construct an excitatory cascade $C^E_M(x)$: a cascade with $C^E_M(0)=0$ and
$C^E_M(\infty)>0$. This is done by piling up a set of excitatory Hill equations $\{\tilde{y}^E_m(x);
\>\> m=1,...,M\}$, as shown in figure \ref{fig:proof.01}\textbf{a}. We can also build an inhibitory
cascade $C^I_M(x)$: a cascade with $C^I_M(0)>0$ and $C^I_M(\infty)=0$ in the same way as the
excitatory one, but now an odd number of inhibitory Hill functions must be used together with an
arbitrary number of excitatory units. For simplicity we have built our inhibitory cascades with the
first Hill equation of the cascade being inhibitory and the remaining ones being excitatory
$\{\tilde{y}^I_1(x), \tilde{y}^E_m(x); \>\> m=2,...,M\}$. One such a cascade is shown in figure
\ref{fig:proof.01}\textbf{b}.

			In both excitatory and inhibitory cascades a problem comes up regarding their upper or lower
limits. The first function in the cascade (we shall be using just excitatory units in this paragraph
without loss of generality) is defined for $x_1\in[0,+\infty)$ and takes values over
$\tilde{y}^E_1\in[0,\tilde{y}^E_1(\infty)<+\infty)$. When feeding $\tilde{y}^E_1(x)$ as an input for
the following unit, this second equation will have an input
$x_2=\tilde{y}^E_1\in[0,\tilde{y}^E_1(\infty)<+\infty)$, meaning that it will take values
$\tilde{y}^E_2\in[0,\tilde{y}^E_2(\tilde{y}^E_1(\infty))<\tilde{y}^E_2(\infty))$. We can see that
the last equation of the cascade will never reach its upper limit because its input does not span
the whole domain of positive real numbers, so if we wanted to build our cascade such that it would
have the $C_M(\infty)$ we desire, we would have to set the upper bound of the last equation of the
cascade to be $\tilde{y}_M(\infty)=\gamma C_M(\infty)$, where $\gamma$ is a positive constant. This
is always possible since $\tilde{y}_M(\infty)$ is allowed to take any positive value we desire. The
same reasoning applies for inhibitory cascades. In figures \ref{fig:proof.01}\textbf{a} and
\ref{fig:proof.01}\textbf{b}, the $\tilde{y}_m(\infty)$ of all equations were corrected such that
the cascades have a maximum of roughly $1$. We note that we have got a relative freedom to choose
most of $\tilde{y}_m(0)$ and $\tilde{y}_m(\infty)$. \\

			Using an inhibitory cascade $C_{M}^{I_1}(x^{I_1}=x)$ and two excitatory ones
$C_{M'}^{E_1}(x^{E_1}=x)$, $C_{M''}^{E_2}(x^{E_2}=C_{M}^{I_1}+C_{M'}^{E_1})$, we can build up an
interesting mathematical device. Let these units be such that:
				\begin{eqnarray}
					x_\theta^{E_1} &<& x_\theta^{I_1}, \nonumber \\ 
					C^{I_1}_{M}(0) &<& x_\theta^{E_2} < C^{I_1}_{M}(0)+C^{E_1}_{M'}(\infty), \nonumber \\ 
					C^{E_1}_{M'}(\infty) &<& x_\theta^{E_2}. 
					\label{eq:proof.14}
				\end{eqnarray}

	\begin{figure*}
		\includegraphics[width= 15 cm]{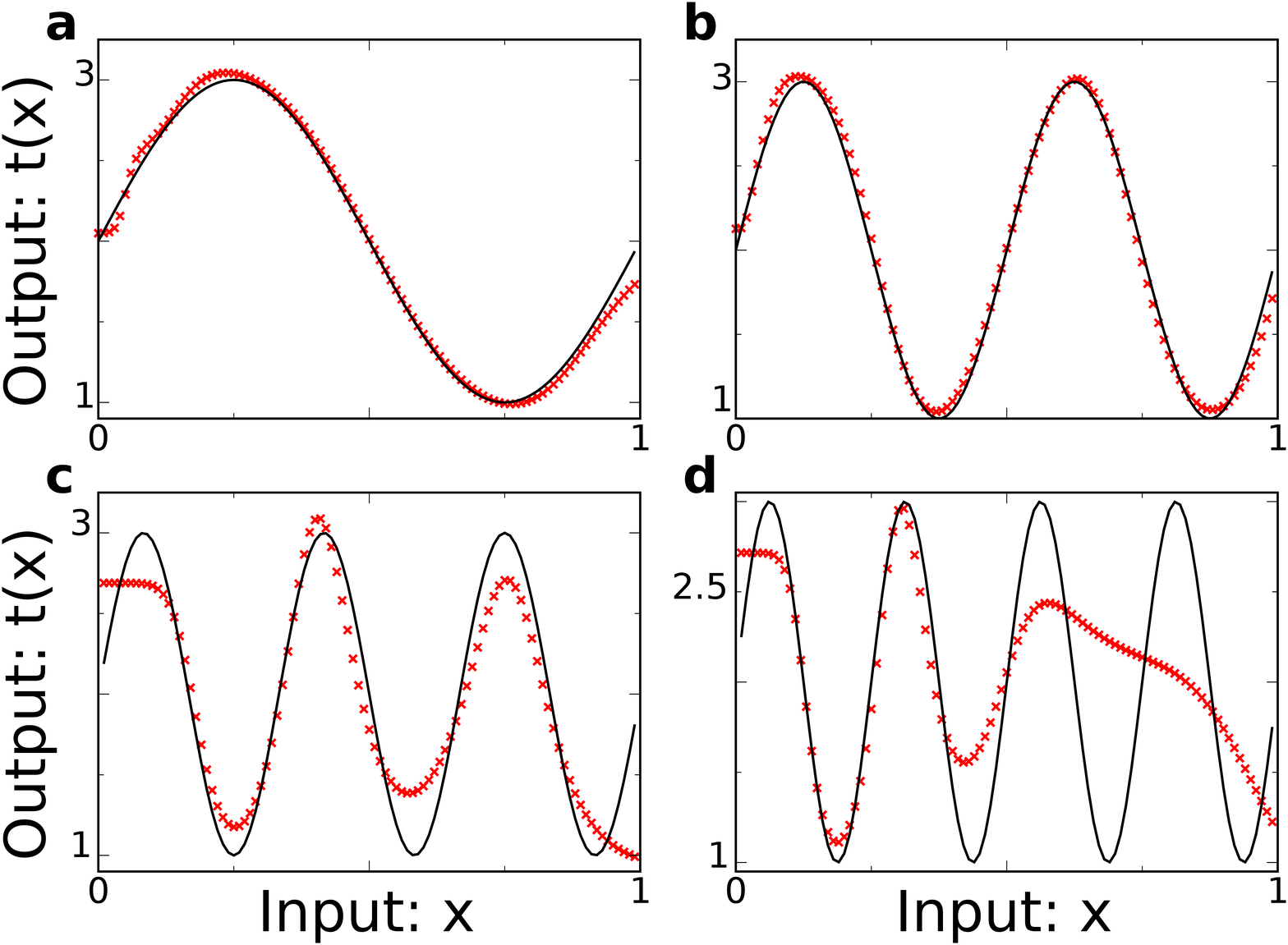}

    \caption{\textbf{A GRN model is trained to fit the function $t(x)=sin(2\pi\nu x)+2$. } A 
three-layered architecture with $4$-$4$-$1$ units at input, hidden and output layers respectively
was chosen. Units at the input layer are regulated by one external promoter whose concentration acts
as input to the whole system and to the target function. Results are shown for different values of
$\nu$: a continuous black curve indicates the actual target function while red crosses indicate the
outcome of the trained GRN at a given input concentration $x$. \textbf{a} $\nu=1$: the easiest task
possible. \textbf{b} $\nu=2$: a slightly more complicated case, but the GRN still performs quite
adequately. \textbf{c} $\nu=3$: as the difficulty increases, the network can not respond as
faithfully as desired anymore. \textbf{d} $\nu=4$: for very complicated instances, two aspects
contribute to distort the output: on the one hand, a larger network might be needed to reproduce
very intricate functions; on the other hand, GRNs are trained using a finite sampling of $t(x)$,
which can only give a hint of the real complexity of the target function.}

		\label{fig:ex.01}
	\end{figure*}

The behavior of a couple of cascades $C^{I_1}_{M}(x)$ and $C^{E_1}_{M'}(x)$ fulfilling these
relationships is plotted in figure \ref{fig:proof.01}\textbf{c}.

			$C_{M''}^{E_2}(x^{E_2})$ is slightly different from all the functions used so far. It would
not, in principle, be exactly described by equation \ref{eq:proof.02} because its input is a sum of
terms, i.e. it would not be regulated by just one promoter. However, using the idea of distributed
computation \cite{Regot-Sole} we could easily envision a case whose input is the sum of two
independent terms while it can still be modeled by the equation that we have so thoroughly
described: $C^{I_1}_{M}(x)$ and $C^{E_1}_{M'}(x)$ must constitute two alternative pathways for the
production of a same protein $X^{E_2}$, such that the concentration of this protein would be the
total yield, as required: $x^{E_2}=C^{I_1}_{M}+C^{E_1}_{M'}$. Assuming that $C^{I_1}_{M}$ and
$C^{E_1}_{M'}$ are exactly this kind of cascades, the total yield of protein $X^{E_2}$ as a function
of $x$ is shown in figure \ref{fig:proof.01}\textbf{d} together with a red line that indicates the
threshold $x_\theta^{E_2}$ of the third cascade.

			Let us note that the threshold of the third cascade $C_{M''}^{E_2}(x^{E_2})$ is reached only
when the two former are activated at the same time, thus serving as an AND logic gate. But the
interest of this object goes beyond the AND gate. Let us call $I(x)\equiv C^{E_2}_{M''}(x)$ to the
exact function implemented by such an assembly of Hill equations as a function of $x$ (figure
\ref{fig:proof.01}\textbf{e}). $I(x)$ can approach a rectangular function $R(x)$ with any desired
accuracy. We just need to adjust $C^{E_2}_{M''}(\infty)$ to the height of $R(x)$ and the
corresponding thresholds $x_\theta^-\equiv x_\theta^{E_1}$ and $x_\theta^+\equiv x_\theta^{I_1}$ to
match those of $R(x)$. A steepen enough transition through the threshold is achieved, as we already
know, by cascading a sufficient number $M$, $M'$, and $M''$ of Hill equations.
			
		\subsection{Feed-forward GRN as Function Approximators}
			\label{subsec:approximators}

			With the elements built so far we are ripe to show that we can approximate any bounded,
positive, continuous function of $N_v$ real, positive variables with a finite feed forward network
of Hill equations.

			For simplicity we proceed for $N_v=1$, without loss of generality. Let there be $f(x)$ a
continuous function taking bounded real, positive values and defined over a finitely bounded subset
of real, positive numbers. We can always find a finite partition $X_\epsilon=\{x_{\epsilon,i};
i=1, ..., \tilde{n}\}$ such that for all $i=1, ..., \tilde{n}-1$ it does not exist any
$x_{\epsilon,i}^+\in[x_{\epsilon,i}, x_{\epsilon,i+1})$ such that
$f(x_{\epsilon,i}^+)>f(x_{\epsilon,i})+\epsilon$ and it does not exits any
$x_{\epsilon,i}^-\in[x_{\epsilon,i}, x_{\epsilon,i+1})$ such that
$f(x_{\epsilon,i}^-)<f(x_{\epsilon,i})-\epsilon$ for any $\epsilon>0$.

			We can find a finite collection of Hill equations with which it is possible to build a set
$I_\epsilon=\{I_{\epsilon,i}(x); \>\> i=1,...,\tilde{n}-1\}$ of functions such that
$I_{\epsilon,i}(x)\simeq f(x_{\epsilon,i})$, if $x\in [x_{\epsilon,i}, x_{\epsilon,i+1})$ and
$I_{\epsilon,i}(x)\simeq 0$ otherwise. Then the function:
				\begin{eqnarray}
					F_\epsilon(x) &=& \sum_{i=1}^{\tilde{n}-1} I_{\epsilon,i}(x) 
					\label{eq:proof.15}
				\end{eqnarray}
approximates $f(x)$ with an error approximately lower than $\epsilon$. 

			To integrate the different outcomes of the collection $I_\epsilon$ into one single output
function $F_{\epsilon}$ it might seem necessary a modeling with several promoters. We would be
facing a case similar to $C^{E_2}_{M''}(x^{E_2})$. But once again we solve this by using the concept
of distributed computation: We must assume that the $I_{\epsilon,i}(x)$ represent different pathways
that synthesize one and the same protein. Each of the $I_{\epsilon,i}(x)$ is responsible of
producing the right output in a narrow domain of $x$, while all the others remain silent. When
looking at our readout, we do not really care which one of the available pathways is producing the
end-product protein, as long as the right concentration -- i.e. a value close enough to $f(x)$ -- is
produced.

			If we allow $f(x)$ to be defined over an unbound set of real, positive numbers then an
infinite number of Hill equations might be needed; but any desired approximation is still
analytically plausible. Also, we can easily extend the analysis to functions of more variables
$N_v>1$ by implementing $I(x_1, ..., x_{N_v})$: the $N_v$-dimensional versions of $I(x)$, which
should be straightforward having distributed computation in mind. And, of course, a vector function
$\overrightarrow{f}(x)$ might be approximated as well by allowing several output Hill equations in
the feed-forward network.

	\section{Back-propagation algorithm for GRN}
		\label{sec:back}
		
		Lacking an algorithm to \emph{train} multilayered artificial neural networks in solving specific
tasks was an important drawback that prevented notable advances in AI for around two decades
\cite{Olazaran}. \emph{Training} a network means finding a set of parameters for its constituents
such that the network as a whole behaves in a desired way. Training -- or learning -- methods are
classified as \emph{supervised} when examples of the expected behavior are provided or
\emph{unsupervised} if a network is expected to work out the structure of a problem on its own.

		The most successful method of supervised learning is the \emph{back-propagation algorithm} \cite
{Rumelhart-Williams} -- a generalization to complex structures of error minimization by hill
climbing that is possible thanks to a nimble handling of partial derivatives and the chain rule.
Notwithstanding this, our networks of Hill equations present several redundancies and
particularities and some choices had to be made to implement a working algorithm. We did so, and the
proof that our choices take on the traininig can be found in section \ref{sec:ex}, where we apply
our methods to four basic machine learning problems. However, discussion remains open about how
correct some of our decisions are. Several alternatives might be equally right, and further research
should be made to tell the most efficient learning strategies for synthetic GRNs. \\

    We will be working with an arbitrary feed-forward network of Hill equations. The network
consists of a series of units, of Hill equations arranged in $N_L$ layers. Each layer is labeled
with the superindex $j$ and has got $N_U^j$ units in it. The first layer will be called the
\emph{imput layer} and the last one will be called the \emph{output layer}. Such a network is a
model for a GRN of promoter proteins that regulate each other in a feed-forward manner such that
proteins regulated at layer $j$ act as promoters of those regulated at layer $j+1$. The
concentration of each unit is governed by a formula similar to equation \ref{eq:proof.01}. We will
rewrite this equation right ahead and therefore we shall adopt a new notation that exhaustively uses
sub- and superindexes. The notation may seem baroque but it is a very appropriate one to derive
back-propagation rules.

		We call $X^j_i$ to the $i$-th protein in the $j$-th layer, and $x^j_i$ to its concentration. We
write the differential Hill equation for the time evolution of this concentration:
			\begin{eqnarray}
				{d x^j_i \over dt} &=& {\alpha^j_{i,0} + \sum_{i'=1}^{N^{j'}_U} \alpha^{jj'}_{ii'} (x^{j'}_{i'})^n 
								\over \beta^j_{i,0} + \sum_{i'=1}^{N^{j'}_U} \beta^{jj'}_{ii'} (x^{j'}_{i'})^n} 
								- \kappa^j_i x^j_i. 
				\label{eq:back.01}
			\end{eqnarray}
Here we have just added a collection of sub- and superindexes to equation \ref{eq:proof.01} and
collapsed the sums within the corresponding symbols, but both equations are essentially the same.
The only actual difference is that we assumed just one and the same Hill coefficient $n$ for all the
proteins in the network. Regarding the notation, we use $j'$ to name the layer whose output works as
input for layer $j$. This usually means $j'=j-1$, but for the input layer we must assume that there
is an extra layer $j'=0$ consisting of the promoters that act as input for the whole system.

		We make this notation even more compact with a classic trick in machine learning: assuming that
there exist several external units with constant concentration $x^{j'}_0=1$. Then:
$\alpha^j_{i,0}\rightarrow\alpha^{jj'}_{i0}$ and $\beta^j_{i,0}\rightarrow\beta^{jj'}_{i0}$. Also,
$\kappa^j_i$ can be integrated into other constants by doing:
$\alpha^{jj'}_{ii'}\rightarrow\tilde{\alpha}^{jj'}_{ii'}=\alpha^{jj'}_{ii'}/\kappa^j_i$. Equation
\ref{eq:back.01} is rewritten as:
			\begin{eqnarray}
				{1 \over \kappa^j_i}{d x^j_i \over dt} &=& {\sum_{i=0}^{N^{j'}_U} \tilde{\alpha}^{jj'}_{ii'} (x^{j'}_{i'})^n 
												\over \sum_{i=0}^{N^{j'}_U} \beta^{jj'}_{ii'} (x^{j'}_{i'})^n} 
												- x^j_i, 
				\label{eq:back.02}
			\end{eqnarray}
such that the concentration of protein $X^j_i$ when the network reaches a steady state is: 
			\begin{eqnarray}
				x^j_i &=& {\sum_{i=0}^{N^{j'}_U} \tilde{\alpha}^{jj'}_{ii'} (x^{j'}_{i'})^n 
							\over \sum_{i=0}^{N^{j'}_U} \beta^{jj'}_{ii'} (x^{j'}_{i'})^n}. 
				\label{eq:back.03}
			\end{eqnarray}

		The set of equations for the whole network successively determines the concentration of proteins
at each layer once the concentration of the promoters of the input layer is given. We read out the
concentration of proteins in the output layer as a sort of \emph{result of the computation} that the
GRN implements. We see that the network computes in a feed-forward manner. At the training phase,
this outcome is compared to some target function that encompasses the behavior that is expected from
the network. This will be clearer in section \ref{sec:ex} when examples are shown. By now it is
enough to assume that an error can be derived by comparing the target function to the result yielded
by the network. This error will be propagated backwards throughout the network, and the parameters
($\tilde{\alpha}^{jj'}_{ii'}$ and $\beta^{jj'}_{ii'}$) will be modified according to how much of the
error each of them is responsible of. To implement this back-propagation concept, the following
derivatives will be necessary:
			\begin{eqnarray}
				{\partial x^j_i \over \partial \tilde{\alpha}^{jj'}_{ik'}} &=& 
					{(x^{j'}_{k'})^n \over \sum_{i=0}^{N^{j'}_U} \beta^{jj'}_{ii'} (x^{j'}_{i'})^n}. \nonumber \\ 
				{\partial x^j_i \over \partial \beta^{jj'}_{ik'}} &=& 
					- x^j_i\cdot{(x^{j'}_{k'})^n \over \sum_{i=0}^{N^{j'}_U} \beta^{jj'}_{ii'} (x^{j'}_{i'})^n}. \nonumber \\ 
				{\partial x^j_i \over \partial x^{j'}_{i'}} &=& 
					{n\over x^{j'}_{k'}}\left[{\tilde{\alpha}^{jj'}_{ik'}\cdot{\partial x^j_i 
						\over \partial \tilde{\alpha}^{jj'}_{ik'}} 
						+ \beta^{jj'}_{ik'}\cdot {\partial x^j_i \over \partial \beta^{jj'}_{ik'}}}\right]. 
				\label{eq:back.04}
			\end{eqnarray}

		The next step is to make use of the error function $\epsilon(x^{N_L}_i; t_i)$ whose existence we
just assumed. This depends on the concentrations $x^{out}\equiv \{x^{N_L}_i\}$ at the output layer
and on a target function $T\equiv\{t_i\}$ that has got as many components as units there are in the
output layer. Let us note that both $x^{out}$ and $T$ are a function of the concentration of the
promoters of the input layer. We define the \emph{partial error}:
			\begin{eqnarray}
				\delta^j_i &=& -{\partial \epsilon(x^{out}; T) \over \partial x^j_i} 
				\label{eq:back.06}
			\end{eqnarray}
of the $i$-th unit in layer $j$. For the output layer the partial error can be obtained straight
away by taking derivatives in the error function. For other layers it is necessary to use the chain
rule as it follows:
			\begin{eqnarray}
				\delta^{j'}_{i'} &=& -{\partial \epsilon(x^{out}; T) \over \partial x^{j'}_{i'}} = 
									\sum_{i=1}^{N^j_U}\left( -{\partial e \over \partial x^j_i}
							\cdot {\partial x^j_i\over \partial x^{j'}_{i'}} \right) = \nonumber \\ 
							&=& \sum_{i=1}^{N^j_U} \delta^j_i \cdot {\partial x^j_i\over \partial x^{j'}_{i'}}. 
				\label{eq:back.07}
			\end{eqnarray}
This allows us to compute the error of units at layer $j'$ once the errors in $j$ are known, thus
\emph{back-}propagation. Knowing these partial errors at each unit it is easy to calculate the
gradient of the error with respect to the different parameters of the network
$\tilde{\alpha}^{jj'}_{ik'}$ and $\beta^{jj'}_{ik'}$ and apply a hill climbing rule on them:
			\begin{eqnarray}
				\Delta \tilde{\alpha}^{jj'}_{ik'} &=& -\eta \cdot {\partial \epsilon(x^{out}; T) 
													\over \partial \tilde{\alpha}^{jj'}_{ik'}} 
										= -\eta\cdot {\partial \epsilon \over \partial x^j_i}\cdot {\partial x^j_i 
													\over \partial \tilde{\alpha}^{jj'}_{ik'}} = \nonumber \\
									 &=& -\eta\delta^j_i{\partial x^j_i \over \partial \tilde{\alpha}^{jj'}_{ik'}}, \\ 
				\Delta \beta^{jj'}_{ik'} &=& -\eta \cdot {\partial \epsilon(x^{out}; T) 
											\over \partial \beta^{jj'}_{ik'}} 
								 = -\eta\cdot {\partial \epsilon \over \partial x^j_i}\cdot {\partial x^j_i 
											\over \partial \beta^{jj'}_{ik'}} = \nonumber \\
								&=& -\eta\delta^j_i{\partial x^j_i \over \partial \beta^{jj'}_{ik'}}; 
				\label{eq:back.08}
			\end{eqnarray}
where $\eta$ represents a learning rate whose value has to be chosen externally for each training
task, and $\tilde{\alpha}^{jj'}_{ik'}\rightarrow \tilde{\alpha}^{jj'}_{ik'}+\Delta
\tilde{\alpha}^{jj'}_{ik'}$ and $\beta^{jj'}_{ik'} \rightarrow \beta^{jj'}_{ik'}+\Delta
\beta^{jj'}_{ik'}$ are the update rules for each parameter at each training step.

		We note straightaway that such an update rule can lead to negative values of
$\tilde{\alpha}^{jj'}_{ik'}$ or $\beta^{jj'}_{ik'}$, which we would like to avoid because these
constants are always positive in biological systems. This issue is solved using the following change
of variables:
			\begin{eqnarray}
				\tilde{\alpha}^{jj'}_{ik'} &\rightarrow& a^{jj'}_{ik'} = +\sqrt{\tilde{\alpha}^{jj'}_{ik'}} \Rightarrow 
												\tilde{\alpha}^{jj'}_{ik'}=(a^{jj'}_{ik'})^2 \Rightarrow \nonumber \\ 
					&\Rightarrow& {\partial x^j_i \over \partial a^{jj'}_{ik'}} 
							= 2 a^{jj'}_{ik'} {\partial x^j_i \over \partial \tilde{\alpha}^{jj'}_{ik'}}, \nonumber \\		
					\beta^{jj'}_{ik'} &\rightarrow& b^{jj'}_{ik'} = +\sqrt{\beta^{jj'}_{ik'}} \Rightarrow 
												\beta^{jj'}_{ik'}=(b^{jj'}_{ik'})^2 \Rightarrow \nonumber \\ 
					&\Rightarrow& {\partial x^j_i \over \partial b^{jj'}_{ik'}} = 2 b^{jj'}_{ik'} {\partial x^j_i \over \partial \beta^{jj'}_{ik'}}; 
				\label{eq:back.05}
			\end{eqnarray}
and implementing update rules on $a^{jj'}_{ik'}$ and $b^{jj'}_{ik'}$ instead. This was made in all
the examples shown. Using a similar strategy we can further impose that $\tilde{\alpha}^{jj'}_{ik'}$
and $\beta^{jj'}_{ik'}$ fall within a realistic range of values, but in this paper we did not go
that far and we did not care that much about the numerical values obtained after training as long as
they were positive.

	\begin{figure}
		\includegraphics[width=0.35\textwidth]{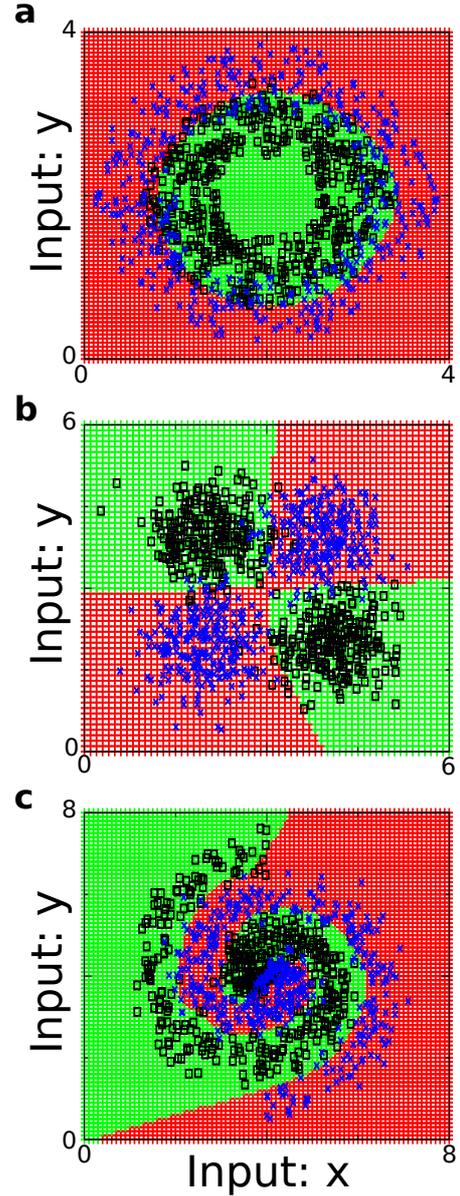}

    \caption{\textbf{Classification problems. } Two inputs
$x$ and $y$ represent the concentrations of the regulating input proteins $X$ and $Y$. They can also
be plotted as two-dimensional coordinates. Given a point $(x,y)$, it was considered that the GRN
classified the coordinate as \emph{class 1} if $x^{out}_1>x^{out}_2$ and as \emph{class 2}
otherwise. \emph{Class 1} points are plotted in red and \emph{class 2} are plotted in green. The
GNRs were tested with the coordinates of a regular grid after their training. Along with the GRN's
outcome for the grid, it is plotted the set of training points: \emph{class 1} samples as blue
crosses and \emph{class 2} samples as black squares. The training set was generated with some noise,
such that the classes were not separable with simple geometrical shapes. \textbf{a} The circle
turned out to be the one requiring a less complex network to be solved, a $4$-$4$-$2$ architecture.
\textbf{b} The XOR needed $5$-$10$-$2$ units in each layer. This seems exaggerated for such a
seemingly simple problem. Presumably a smaller network could be reached if additional wiring-
optimization algorithms were used. \textbf{c} The spiral pattern is a difficult one and it required
$10$-$10$-$2$ units to achieve the shown results.}

		\label{fig:ex.02}
	\end{figure}

		With this we are ready to go ahead with some easy machine learning examples. The 
back-propagation rules derived here are just a straightforward implementation of the standard
algorithm applied to ANNs, with the only difference that each pair of units is linked by two
weights. Apart from this, more modern implementations could be done that incorporate dynamic tuning
of $\eta$ or memory effects from the training history, or that exploit some stochasticity to avoid
converging towards local minimums. Such improvements would speed up the training or guarantee that
the desired behavior would be better reproduced. We are not concerned with these issues now: this
work intends to demonstrate the concept of machine learning implemented on models of GRNs. Because
of this, it will be found that examples from the following section require an exaggerated training
period (for nowadays machine learning standards), or that convergence could be better than shown. It
should be kept in mind that there is still plenty of room for improving the presented methods.

	\section{Four easy Machine Learning tasks}
		\label{sec:ex}
		
		As a proof of concept, we used feed-forward networks of Hill equations and the proposed training
rules to solve four easy machine learning problems: one fitting and three classification tasks.

		During the learning phase we appreciated that our algorithm is quite sensitive to initial
conditions. The initial values of $\tilde{\alpha}^{jj'}_{ii'}$ and $\beta^{jj'}_{ii'}$ are chosen
randomly before learning, and an unlucky initialization can harm convergence towards the desired
network, probably because the algorithm gets stuck in local minimums. This is common also in very
basic ANNs implementations.

		We were not specially interested in constructing optimal GRNs (meaning GRNs with less units or
connections and yet with a good performance). Network growing or pruning algorithms could be
implemented for this end. Our main interest for this paper was to show some working examples of the
ideas developed, and thus we proceeded using the most basic methods.

		Because some freedom remained to set up the $\kappa^j_i$ and $n$ we chose $n=2$, which is a very
realistic value for systems modeled by Hill equations, and $\kappa=1$. This applies to all the
examples following.

	\begin{figure}
		\includegraphics[width=0.4\textwidth]{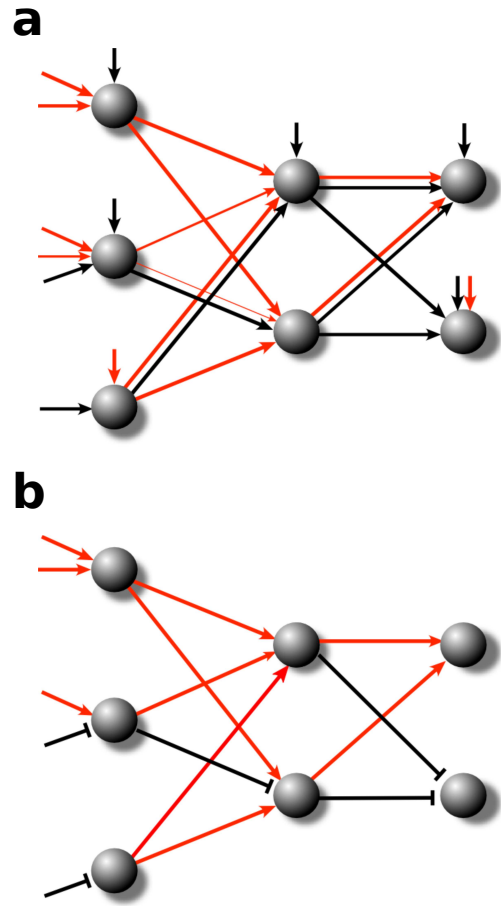}

    \caption{\textbf{Synthetic GRN designed with the backpropagation algorithm.} \textbf{a} Raw
representation of the network obtained for the circle classification task: $4$-$4$-$2$ units are
used in the input, hidden and output layers respectively. Genes in the input layer are regulated by
two input promoters. Each pair of nodes is linked by two arrows, one representing the value of
$\alpha^{jj'}_{ii'}$ (red) and the other one representing $\beta^{jj'}_{ii'}$ (black). The basal
expression is mediated by $\alpha^{jj'}_{i0}$ and $\beta^{jj'}_{i0}$, encoded by the vertical
arrows. The width of each arrow bears some information about how strong each affinity term is. A
connection will have inhibitory or excitatory effects depending on the ratio
$\alpha^{jj'}_{ii'}/\beta^{jj'}_{ii'}$. \textbf{b} Schematic representation of the same network: the
relationships between pairs of nodes have been reduced to excitatory (red) and inhibitory (black).
The outcome of our algorithmic design of GRNs can be compared with the reconstruction of real webs
of interacting genes from figure \ref{fig:intro.01}. The parallelisms suggest that the cell does
implement computations in a not so different manner.}

	\label{fig:ex.03}
	\end{figure}

		\subsection{Using GRNs to fit a function} 
			\label{subsec:fit}
			
			In the first task we wanted our network to behave such that when an input $x$ was given to the
input layer, then the output layer would yield $t(x)$, where $t(x)$ is an arbitrary mathematical
function that we shall call the \emph{target} of our training. In biological terms, the proteins at
the input layer of our GRN would be exposed to a concentration $x$ of their promoter protein $X$,
and it would be wished that the concentration $x^{out}$ of the protein $X^{out}$ at the output layer
would be exactly $t(x)$.

			We used $t(x)=sin(2\pi\nu x)+2$ with $\nu=1$, $2$, $3$, and $4$ and $x\in[0,1)$. This is a
function of only one variable, thus proteins in the input layer would be controlled by just one
external promoter; it is always positive, as required in section \ref{sec:proof} for minimally
realistic systems; and it is unidimensional, thus one only unit is required in the output layer.

			Many different architectures were tried, but the results shown here correspond to networks
with three layers: the input one with 4 units, a hidden layer with 4 more units, and the output
layer. We used the error function:
				\begin{eqnarray}
					\epsilon(x^{out};T) &=& {\left(x^{out}-t(x)+\xi\right)^2\over 2}, 
					\label{eq:ex.01}
				\end{eqnarray}
where $\xi$ is a Gausian noise with $0$ mean and standard deviation $0.1$, so that perfect examples
did probably not show up

			For the training, $100$ data points $(x,t(x))$ were generated and each point was presented
$1000$ times to the GRN. The back-propagation algorithm was applied right after each data
presentation. Results for the different $\nu$ are shown in figure \ref{fig:ex.01}. We appreciate how
the performance gets worst as the complexity of the function within $[0,1)$ increases. Apart from
the obvious need for more units to reconstruct finer details, we must concede that the training
points do not need to reveal the whole structure of the target function because they are just a
noisy, finite sample of it.

		\subsection{Three classification tasks}
			\label{subsec:class}

			We made up three classification tasks of varying difficulty. For each task we generated $1000$
data points belonging to classes $1$ or $2$. Each point consists of a two-dimensional coordinate and
a label is attached that indicates its class. The different classes are arranged on a plane. We
identify each task as \emph{circle}, \emph{XOR}, and \emph{spiral} after the geometric disposition
of the data (see figure \ref{fig:ex.02}). Noise was added such that the classes overlap and they are
not easily separable. These points were used to train the network: back-propagation was applied
right after the presentation of each data point and each point was presented $100$ times. When the
learning phase was over, a grid that covered the area spanned by the training samples was given to
the network to test its performance.

			The input now is a two-dimensional coordinate, thus proteins of the input layer have got two
external promoters $X$ and $Y$ with concentrations $x$ and $y$. It would have been possible to read
out the result with just one output unit, attending to whether its concentration was above or below
an arbitrary threshold. It was decided, though, to always use two units $X^{out}_1$ and $X^{out}_2$
in the output layer and interpret the result as \emph{class 1} if $x^{out}_1>x^{out}_2$ and as
\emph{class 2} otherwise. It was used the following error function:
				\begin{eqnarray}
					\epsilon(x^{out};T) &=& { (x^{out}_1-t_1)^2 + (x^{out}_2-t_2)^2 \over 2}, 
					\label{eq:ex.02}
				\end{eqnarray}
where $T=(1,0)$ if the input belonged to class $1$ and $T=(0,1)$ if it belonged to class $2$. 

			The circle task (figure \ref{fig:ex.02}\textbf{a}) happened to be the less demanding one: it
was solved with just $3$ units in the input layer and $2$ units in the hidden layer, plus two output
units as said before. The network obtained is represented in figure \ref{fig:ex.03}. This reminds us
of the small feed-forward network motifs from real cells (figure \ref{fig:intro.01}).

			Because the tasks were of growing difficulty, the number of units needed to solve each problem
changed. The XOR (figure \ref{fig:ex.02}\textbf{b}) required $5$ and $10$ units in the input and
hidden layers, and the spiral (figure \ref{fig:ex.02}\textbf{c}) could not be solved with less than
$10$ units in each non-output layer. Both of them had $2$ units as an output.

			Let us recall once more that it was not sought any optimization in terms of wiring or number
of nodes. It can be expected that much smaller GRNs can be designed to solve either of these
problems (including also the function fitting task) if network growing or pruning algorithms were
employed. Such methods are quite common in ANNs architecture optimization, which is a non-trivial
problem. Taking into account that we did not care about the wiring that much, the results obtained
are very satisfactory.

	\section{Discussion}
		\label{sec:disc}
		
		Within the developing framework of biological computation, in this paper we proposed that
feed-forward GRNs are a suitable substrate to implement the basics of machine learning. They can
cope perfectly with solving regression problems or with data classification, as it has been
explicitly shown in four examples. This opens the door to more complicated AI applications such as
data mining, data series prediction, clustering, signal filtering, etc. Of course, using networks of
Hill equations does not suppose an improvement in performance when compared to existing ANNs. This
was not the aim of this work. Our purpose was rather to show that realistic biological systems have
got a complexity enough as to solve machine learning tasks. We did so with a constructive -- enough,
but not completely rigorous -- mathematical proof that the chosen GRN model can approximate any
bounded, positive, continuous function and presenting four practical examples.

		GRNs appear as a quite adequate means to implement ANNs-like architectures. Already in cellular
systems, genes appear arranged in networks and execute very complex duties. Because of this network
disposition, the models we work with are readily suitable to derive error back-propagation rules.
This is the most successful algorithm for training ANNs in implementing arbitrary functions. It is
an analytical tool, a theoretical device; so knowing these rules for networks of Hill equations we
can design ideal GRNs that would behave as we wished.

		Of course, back-propagation does not provide a technique to build synthetic networks in vivo,
but rather the instructions for what will be needed when the moment arrives. With the idea of
providing a faithful, realistic plan, there are a series of constrains that we should sooner or
later impose in the large set of parameters that our GRNs would have. For the present paper we just
demanded that our networks should always have positive affinities and positive basal expression
levels, and that Hill coefficients should be low enough. For the practical examples we used $n=2$,
which is among the lowest possible. Also, in these examples we did not care that much if the values
of certain parameters turned out exaggeratedly huge or small after training -- as it was the case,
among other reasons because we have been working with equations in arbitrary dimensions. But
requesting that these parameters fall within a physically plausible range and many other necessary
constraints can be easily incorporated into the back-propagation rules so that we would arrive to
feasible GRNs.

		An immediate critique to the idea introduced in this paper is that its development in actual
biological systems may require a very fine-tuning of parameters that is out of reach for 
state-of-the-art experimental techniques. The scenario might be even worst: since stochasticity is
so in the core of molecular biology, narrowly controlling gene-gene interactions might never be
possible. But this observation is not so straightforwardly true. Indeed, a large GRN means tens or
hundreds of parameters to be tuned. They could just be flexible enough as to carry out a desired
task with several different sets of parameters, or with parameters within certain ranges, thus
dismissing the need for fine-tuning. ANNs implement computations in a distributed and emergent
manner, usually avoiding that very specific pieces are responsible of very precise parts of the
calculi. This makes ANNs both flexible and robust against stochastic events, and it is not strange
that different ANNs work in a similar and stable way while having very different internal settings.
These features are very interesting for synthetic circuits design.

		It might not only be the case that fine-tuning is not required. Cutting-edge experimental
techniques \cite{Wang-Church} could allow us to treat whole genomes as black boxes upon which
accelerated evolution aided	by artificial selection could implement training algorithms similar in
essence to a Hebbian learning. In such a case we would not need to know any actual numerical values
for parameters that we would need to write down into a genome. Rather, the evolving system could
self-organize to implement any function that we wished, as long as it would have enough
computational complexity.

		As already outlined in section \ref{sec:back}, our algorithm implementation required some
assumptions. For example, we have chosen to use the constant Hill coefficient $n=2$, or to consider
both $\tilde{\alpha}^{jj'}_{ii'}$ and $\beta^{jj'}_{ii'}$ as the weights between pairs of units:
this may enclose some redundancy. We have also used the most basic learning algorithm possible,
while nowadays several techniques exist that make the training of a network far more efficient,
meaning better performance and reduced learning times. This very basic algorithm was already enough
for the proof of concept that we intended. However, all this work should be revised some day if the
design of actual GRNs were sought: we should find the most economic networks, thus rewiring or
pruning techniques would be a nice extension; and those networks working in parameter ranges in
which less fine-tuning is required would also be preferred. Besides, designing networks that are
robust to noise or that somehow incorporate it as a computing feature would be desirable, since
stochasticity pervades real regulatory systems. All these questions can be addressed numerically and
analytically using some of the methods developed in this paper, or incorporating to them whatever
extensions were needed. Finally, similar algorithms can be derived for recurrent and dynamical
models of GRNs. This would hopefully reveal an even greater power of networks of genes as
computational devices.

\section*{Acknowledgements}

We would like to thank the members of the Complex Systems Lab. This work has been supported by
grants from the Spanish MINECO, a European Research Council
Advanced Grant, the Botin Foundation and by the Santa Fe Institute.


\begin{thebibliography}{} 

		\bibitem[Alon 2006]{bookAlon}
			Alon U (2006)
			An Introduction to Systems Biology: Design Principles of Biological Circuits. 
			Taylor \& Francis. 
		
		\bibitem[Alon 2007]{Alon}	
			Alon U (2007)
			Network motifs: theory and experimental approaches. 
			Nature 8:450-61 

		\bibitem[Amos 2004]{Amos2004}
			Amos M (2004)
			Cellular Computing. 
			Oxford University Press

		\bibitem[Amos 2008]{Amos2008}
			Amos M (2008)
			Genesis Machines: The New Science of Biocomputing. 
			MIT Press, Cambridge MA
			
		\bibitem[Auslander et al. 2012]{Auslander-Fussenegger}
			Ausl\"ander S, Ausl\"ander D, M\"uller M, Wieland M, and Fussenegger M (2012) 
			Programmable single-cell mammalian biocomputers. 
			Nature 487(7405):123-127

		\bibitem[Bashor et al. 2010]{Bashor-Lim}
			Bashor CJ, Horwitz AA, Peisajovich SG, and Lim WA (2010)
			Rewiring cells: synthetic biology as a tool to interrogate the organizational principles of living systems. 
			Annu Rev Biophys 39:515-537
			
		\bibitem[Benenson 2012]{Benenson}
			Benenson Y (2012)
			Biomolecular computing systems: principles, progress and potential. 
			Nature Rev Genet 13:455-468

		\bibitem[Bishop 2006]{Bishop}
			Bishop CM (2006)
			Pattern Recognition and Machine Learning. 
			Springer-Verlag, New York
			
		\bibitem[Bratsun et al. 2005]{Bratsun-Hasty} 
			Bratsun D, Volfson D, Tsimring LS, and Hasty J (2005)
			Delay-induced stochastic oscillations in gene regulation. 
			Proc Natl Ac Sci 102(41):14593-14598
			
		\bibitem[Bray 1990]{Bray1990}
			Bray D (1990) 
			Intracellular Signaling as a Parallel Distributed Process. 
			J theor Biol 143:215-231 
			
		\bibitem[Bray 1995]{Bray1995}
			Bray D (1995)
			Protein molecules as computational elements in living cells. 
			Nature 376:307-312

		\bibitem[Brenner 2012]{Brenner}
			Brenner S (2012) 
			Life's code script. 
			Nature 482:461 

			
		\bibitem[Dayan and Abbott 2005]{Dayan-Abbott}
			Dayan P and Abbott LE (2005) 
			Theoretical Neuroscience. 
			MIT Press, Cambridge MA
			
		\bibitem[Gardner et al. 2000]{Gardner-Collins}
			Gardner TS, Cantor CR, and Collins JJ (2000)
			Construction of a genetic toggle switch in \emph{Escherichia coli}. 
			Nature 403:339-342
			
		\bibitem[Goutelle 2008]{Goutelle-Maire}
			Goutelle S, Maurin M, Rougier F, Barbaut X, Bourguigon L, Ducher M, and Maire P (2008) 
			The Hill equation: a review of its capabilities in pharmacological modelling. 
			Fund Clin Pharmacol 22:633-636
			
		\bibitem[Hasty et al. 2001]{Hasty-Collins}
			Hasty J, Isaacs F, Dolnik M, McMillen D, and Collins JJ (2001) 
			Designer gene networks: Towards fundamental cellular control. 
			Chaos 11(1):207-220

		\bibitem[Hill 1910]{Hill}
			Hill AV (1910)
			The possible effects of the aggregation of the molecules of h\ae moglobin on its dissociation curves. 
			J Physiol 40:iv-vii
			
		\bibitem[Hoffman-Sommer et al. 2012]{Hoffman-Kipp}
			Hoffman-Sommer M, Supady A, and Klipp E (2012)
			Cell-to-cell communication circuits: quantitative analysis of synthetic logic gates. 
			Front Physio 3:287 


		\bibitem[Jacob and Monod 1961]{Jason-Monod}
			Jacob F and Monod J (1961) 
			Genetic regulatory mechanisms in the synthesis of proteins. 
			J Mol Biol 3:318-356

		\bibitem[Kauffman 1993]{Kauffman}
			Kauffman SA (1993)
			Origins of Order. 
			Oxford University Press
			
		\bibitem[Kawczy\'nski and Legawiec 2001]{Kawczynski-Legawiec} 
			Kawczy\'nski AL, Legawiec B (2001)
			Two-dimensional model of a reaction-diffusion system as a typewriter. 
			Phys Rev E 54:056202-1

		\bibitem[Kitano 2001]{Kitano}
			Kitano H (2001) 
			oundations of Systems Biology. 
			MIT Press, Cambridge MA

		\bibitem[Lim 2010]{Lim}
			Lim WA (2010)
			Designing customized cell signalling circuits. 
			Nat Rev Mol Cell Bio 11:393-403

		\bibitem[Mac\'ia et al. 2009a]{Macia-Sole2009a}
			Mac\'ia J, Widder S, and Sol\'e RV (2009)
			Specialized or flexible feed-forward loop motifs: a question of topology. 
			BMC Syst Biol 3:84

		\bibitem[Mac\'ia et al. 2009b]{Macia-Sole2009b}
			Mac\'ia J, Widder S, and Sol\'e RV (2009)
			Why are cellular switches Boolean? General conditions for multistable genetic circuits. 
			J Theor Biol 261(1):126-135

		\bibitem[Mac\'ia et al. 2012]{Macia-Sole2012}
			Mac\'ia J, Posas F, and Sol\'e RV (2012)
			Distributed computation: the new wave of synthetic biology devices. 
			Trends Biotechnol 30(6):342-349

		\bibitem[Maass et al. 2002]{Maass-Markram}
			Maass W, Natschlaeger T and Markream H (2002)
			Real-time computing without stable states: A new framework for neural computation based on perturbations.
			Neural Comput 14(11):2531-2560

		\bibitem[Mattiussi and Floreano 2007]{Mattiussi-Floreano}
			Mattiussi C and Floreano D (2007)
			Analog genetic encoding for the evolution of circuits and networks. 
			IEEE T Evolut Comput 11:596-607
			
		\bibitem[McCulloch and Pitts 1943]{McCulloch-Pitts}
			McCulloch WS and Pitts WH (1943)
			A logical calculus of the ideas immanent in nervous activity. 
			B Math Biophys 5:115-133
		
		\bibitem[Mondrag\'on et al. 2011]{Mondragon-Hasty} 
			Mondrag\'on-Palomino O, Danino T, Slimkhanov J, Tsimring L, and Hasty J (2011)
			Entrainment of a Population of Synthetic Genetic Oscillators. 
			Science 333:1315-1319
			
		\bibitem[Nurse 2008]{Nurse}
			Nurse P (2008)
			Life, logic and information. 
			Nature 454:424-426

		\bibitem[Olazaran 1989]{Olazaran}
			Olazaran M (1989) 
			A Sociological Study of the Official History of the Perceptrons Controversy. 
			Soc Stud Sci 26(3):611-659
			
		\bibitem[Regot et al. 2010]{Regot-Sole} 
			Regot S, Mac\'ia J, Conde N, Furukawa K, Kjell\'en J, Peeters T, Hohmann S, de Nadal E, Posas F, and Sol\'e R (2010)
			Distributed biological computation with multicellular engineered networks. 
			Nature 469:207-211
			
			
		\bibitem[Rodriguez-Caso et al. 2005]{CRCaso} 
			Rodriguez Caso C, Medina MA, Sol\'e RV (2005) 
			Topology, tinkering and evolution of the human transcription factor network.  
			FEBS J. 272: 6423-6434.		
			
			
		\bibitem[Rothemund et al. 2004]{Rothemund-Winfree} 
			Rothemund PWK, Papadakis N, Winfree E (2004) 
			Algorithmic Self-Assembly of DNA Sierpinski Triangles. 
			PLoS Biol 2(12):e424

		\bibitem[Rumelhart et al. 1986]{Rumelhart-Williams}
			Rumelhart DE, Geoffrey EH, and Williams RJ (1986)
			Learning representations by back-propagating errors. 
			
		\bibitem[Saez-Rodriguez et al. 2009]{SaezRodriguez-Sorger}
			Saez-Rodriguez J, Alexopoulos LG, Epperlein J, Samaga R, Lauffenburger DA, Klamt S, and Sorger PK (2009)
			Discrete logic modelling as a means to link protein signalling networks with functional analysis of mammalian signal transduction. 
			Mol Syst Biol 5:331 
	
		\bibitem[Santill\'an 2008]{Santillan}
			Santill\'an M (2008)
			On the Use of Hill Functions in Mathematical Models of Gene Regulatory Networks. 
			Mat Model Nat Phenom 3(2):85-97
			
		\bibitem[S\^irbu et al. 2011 and references therein]{Sirbu-Crane}
			S\^irbu A, Ruskin HJ, and Crane M (2012)
			Stages of Gene Regulatory Network Inference: the Evolutionary Algorithm Role

		\bibitem[Sol\'e and Mac\'ia 2013]{Sole-Macia}
			Sol\'e RV and Mac\'ia J (2013) 
			Expanding the landscape of biological computation with synthetic multicellular consortia. 
			Nat Comp 10.1007/s11047-013-9380-y
			
		\bibitem[Tabor et al. 2009]{Tabor-Ellington}
			Tabor JJ, Salis H, Simpson ZB, Chevalier AA, Levskaya A, Marcotte EM, Voigt CA, and Ellington AD (2009)
			A Synthetic Genetic Edge Detection Program.
			Cell 137(7):1272-1281

		\bibitem[Wang et al. 2009]{Wang-Church}
			Wang HH, Isaacs FJ, Carr PA, Sun ZZ, Xu G, Forest CR, and Church GM (2009) 
			Programming cells by multiplex genome engineering and accelerated evolution. 
			Nature 460:894-899
			
		\bibitem[Widder et al. 2012]{Widder-Macia}
			Widder S, Sol\'e R, and Mac\'ia J (2012)
			Evolvability of feed-forward loop architecture biases its abundance in transcription networks. 
			BMC Syst Biol 6:7
			Nature 323:533-536

	\end{thebibliography}
\end{document}